\documentclass[preprint,aps,showpacs,superscriptaddress,tightenlines,a4paper]{revtex4}

\usepackage{amssymb}
\usepackage{amsmath}
\usepackage{bbold}
\usepackage{mathrsfs}
\usepackage{paralist}
\usepackage{graphicx}
\usepackage{xcolor}

\newcommand{\ii}{{\rm i}}
\newcommand{\e}{{\rm e}}

\begin{document}
%
%
    \title{Non--reflecting permittivity profiles and the spatial Kramers--Kronig relations}
%
%
    \author{S. A. R. Horsley}
    \affiliation{Department of Physics and Astronomy, University of Exeter,
Stocker Road, Exeter, EX4 4QL}
    \email{s.horsley@exeter.ac.uk}
    \author{M. Artoni}
    \affiliation{European Laboratory for Nonlinear Spectroscopy, Sesto Fiorentino, Italy}
    \affiliation{Department of Engineering and Information Technology CNR-IDASC Sensor Lab, Brescia University, Brescia, Italy}
    \author{G. C. La Rocca}
    \affiliation{Scuola Normale Superiore and CNISM, Pisa, Italy}
%
%
    \begin{abstract}
    We show that if the permittivity profile of a planar dielectric medium is an analytic function in the upper (lower) half complex \emph{position} plane then it won't reflect radiation incident from the left (right), whatever the angle of incidence.  Consequently, using the spatial Kramers--Kronig relations one can derive a real part of a permittivity profile from some given imaginary part (or vice versa), such that the reflection is guaranteed to be zero.  This result is valid for both scalar and vector wave theories, and may have relevance for efficiently absorbing radiation, or reducing the reflection from bodies.
    \end{abstract}
%
%
    \maketitle
%
%
    A wave propagating through an inhomogeneous medium is almost always reflected to some degree.  This is often practically undesirable, but it is well known that in the case of an abrupt jump in the material parameters the reflection can be suppressed through applying an anti--reflection coating~\cite{macleod2001}.  However, less seems to be understood about what is required for a generic inhomogeneous medium not to reflect any radiation.  Having said this, there are some famous examples of non--reflecting material profiles. One long--known example is the hyperbolic secant profile, which can be found in Landau and Lifshitz~\cite{volume3} and has been very clearly discussed by Lekner~\cite{lekner2007} (see~\cite{thekkekara2014} for an experimental realisation).  More recently the design technique of transformation optics~\cite{pendry2006,leonhardt2006} has been a significant development, giving us a recipe for finding inhomogeneous, anisotropic materials (transformation media) that reflect no radiation whatever the incident field~\cite{pendry2006,valentine2009,pendry2000}.  In the same vein \emph{perfectly matched layers}~\cite{berenger1993}, are a known family of anisotropic lossy media that are closely connected to transformation media, and absorb a wave without producing any reflection~\cite{teixeira1999,popa2011,odabasi2011,sainath2014}.  Another property of inhomogeneous media that can give an absence of reflection is PT--symmetry~\cite{kottos2010,longhi2010,makris2011}.  This is a symmetry where the real and imaginary parts of the permittivity are engineered such that they are invariant under a simultaneous inversion of space and reversal of time.  For complex permittivities this requires regions of gain (\({\rm Im}[\epsilon(x)]<0\)) as well as loss (\({\rm Im}[\epsilon(x)]>0\)).  PT--symmetry guarantees zero reflection in some cases~\cite{lin2011,regensburger2012}, and has been found to be related to the use of complex coordinates in transformation optics~\cite{castaldi2013}.  Metamaterials allow for the realisation of such inhomogeneous permittivity and permeability profiles~\cite{cai2010} through the use of specially designed sub--wavelength elements, and this may allow for the exploration of these new methods for suppressing the reflection of waves.  In particular recent work on so--called `\emph{dispersion engineering}'~\cite{ye2013} has seen the simultaneous control of the real and imaginary parts of the permittivity and permeability which is necessary for the implementation of the following theory.
    \par
    Here we investigate the general problem of finding \emph{isotropic} permittivity profiles with a combination of real and imaginary parts such that the reflection is zero.  For planar media we find the very general condition that when the profile is an analytic function in the upper or lower half complex position plane, and therefore obeys the Kramers--Kronig relations in space, the reflection from respectively the left or from the right vanishes, whatever the angle of incidence.  We note at the outset that this condition is only sufficient and not necessary for zero reflection.
As a corollary of our finding, if the real part of such a non-reflecting permittivity profile is symmetric about some point in space then the corresponding imaginary part always turns out to be antisymmetric about this point, thus exhibiting PT--symmetry.  Therefore one aspect of this work is that, similar to the findings of Castaldi \emph{et. al.}, it also points to a relationship between the use of complex coordinates, the absence of reflection, and PT--symmetry.  
    \begin{figure}[h!]
        \begin{center}
	\includegraphics[width=15cm]{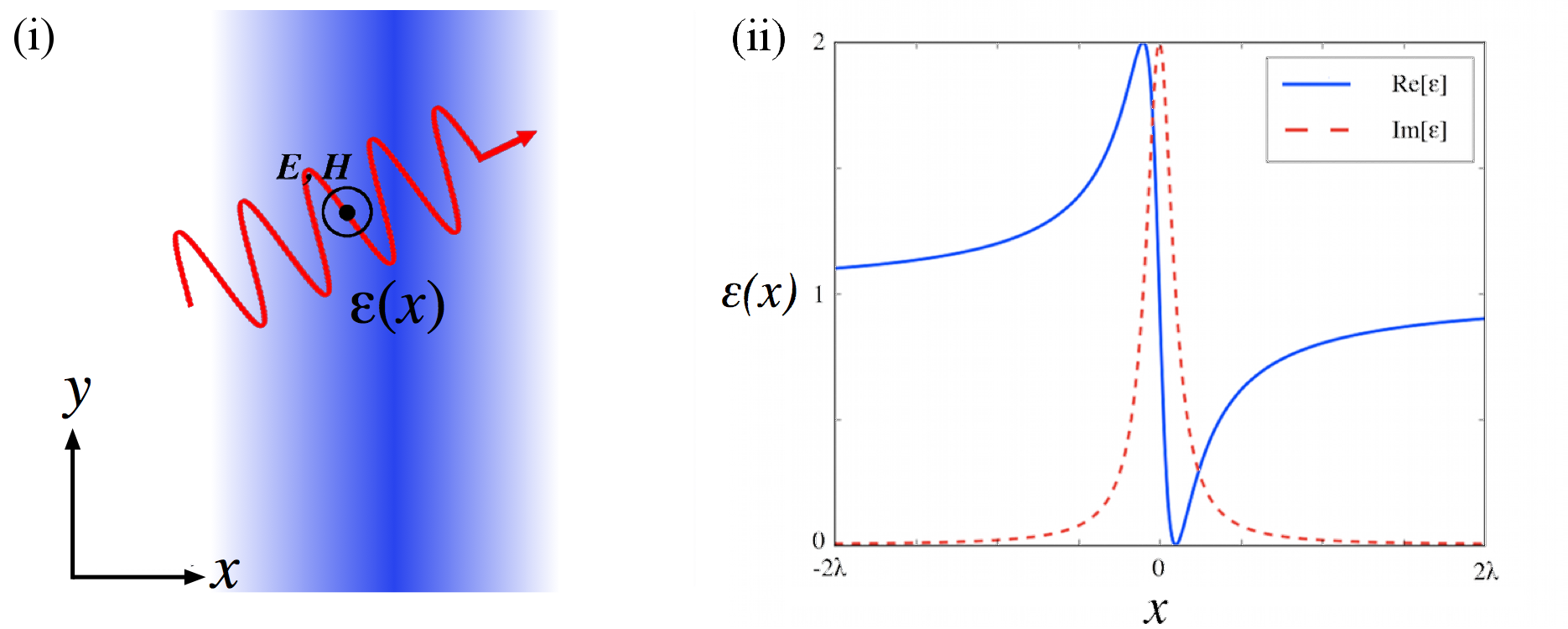}
        \caption{(i) A wave propagates in the \(x\)--\(y\) plane in a medium characterized by an inhomogeneous permittivity \(\epsilon(x)\) (inhomogeneity indicated by the blue shading). (ii) Real (blue) and imaginary (red) parts of the permittivity profile \(\epsilon(x)\) given by equation (\ref{eps_example}) for the parameters \(A=2.0\) and \(\xi=0.1\lambda\) with \(\lambda=2\pi/k_{0}\).\label{eps_figure}}
        \end{center}
    \end{figure}
    \par
    Consider a monochromatic electromagnetic wave propagating in the \(x\)--\(y\) plane within a medium with an inhomogeneous permittivity \(\epsilon(x)\) that tends to a constant positive value \(\epsilon_{b}\) as \(x\to\pm\infty\).  The magnetic permeability is unity \(\mu=1\).  A schematic of this situation is shown in figure~\ref{eps_figure}(i).  The two polarizations are TE (electric field along \(z\)) and TM (magnetic field along \(z\)).  For the TE polarization we can write the electric field as
    \begin{equation}
        E_{z}(x,y)=e_{z}(x)\e^{\ii k_{y}y}
    \end{equation}
    and the \(x\)--dependent amplitude \(e_{z}\) obeys the 1D Helmholtz equation
    \begin{equation}
        \left[\frac{d^{2}}{d x^{2}}+K^{2}+k_{0}^{2}\alpha(x)\right]e_{z}(x)=0.\label{1dwv}
    \end{equation}
    In the above equation the permittivity has the assumed form of the positive background contribution \(\epsilon_{b}\) plus a spatially varying part
    \begin{equation}
        \epsilon(x)=\epsilon_{b}+\alpha(x),\label{eps}
    \end{equation}
    and the wave--number \(K\) is
    \begin{equation}
        K=\sqrt{\epsilon_{b}k_{0}^{2}-k_{y}^{2}}
    \end{equation}
    with \(k_{0}=\omega/c\).  The spatially varying part of the permittivity \(\alpha(x)\) vanishes at large distances from the origin, where the field is made up of plane waves \(\exp(\pm \ii K x)\).
    \par
    Now suppose that we have a right--going wave that comes from infinity \(x=-\infty\) and is incident onto the inhomogeneous permittivity profile.  The effect of the permittivity profile is to produce a scattered field \(e_{s}\), and we can write the total field as
    \begin{equation}
        e_{z}(x)=E_{0}\e^{\ii K x}+e_{s}(x).\label{field-form}
    \end{equation}
    where \(K>0\).  Inserting (\ref{field-form}) into (\ref{1dwv}) we find the inhomogeneous differential equation that governs the scattered field
    \begin{equation}
        \left[\frac{d^{2}}{d x^{2}}+K^{2}+k_{0}^{2}\alpha(x)\right]e_{s}(x)=-k_{0}^{2}\alpha(x)E_{0}\e^{\ii K x}.\label{eseq}
    \end{equation}
    One well--known way to solve equation (\ref{eseq}) is to expand \(e_{s}\) as a series
    \begin{equation}
        e_{s}(x)=\sum_{n=1}^{\infty}e_{s}^{(n)}(x)\label{series}
    \end{equation}
    where the \(n^{\text{th}}\) term is proportional to the \(n^{\text{th}}\) power of \(\alpha\).  The first term in this series---known as the Born approximation in scattering calculations---can be found immediately and is
    \begin{equation}
        e_{s}^{(1)}(x)=-E_{0}k_{0}^{2}\int\frac{dk}{2\pi}G(k)\tilde{\alpha}(k-K)\e^{\ii k x}\label{e1},
    \end{equation}
    where \(\tilde{\alpha}\) is the spatial Fourier transform of \(\alpha(x)\) and \(G(k)\) is the retarded Green function
    \begin{equation}
        G(k)=\frac{1}{(K+\ii\eta)^{2}-k^{2}}
    \end{equation}
    \begin{figure}[h!]
        \begin{center}
	\includegraphics[width=12cm]{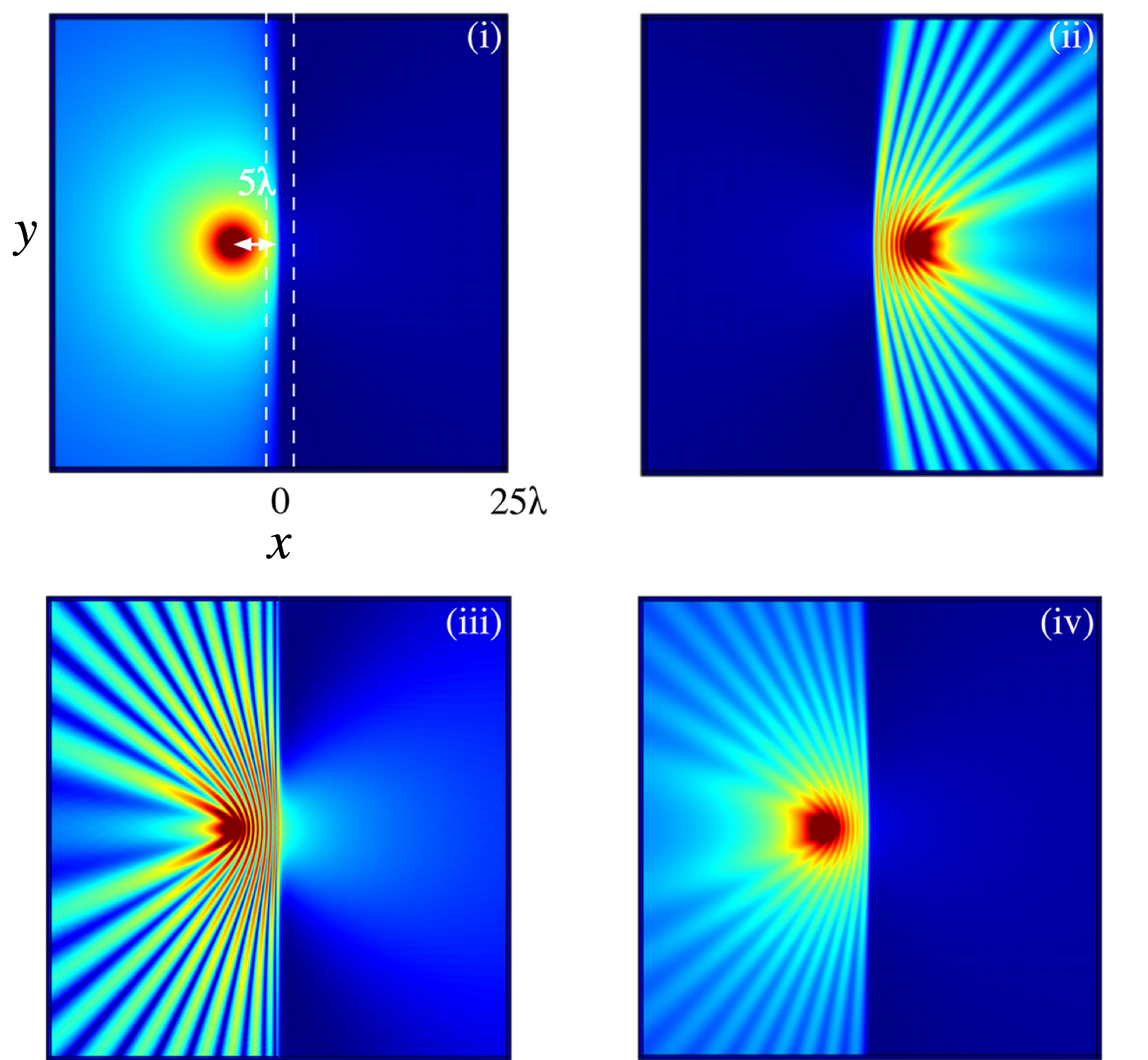}
        \caption{Simulation output generated using Comsol Multiphysics.  Panels (i) and (ii) show the absolute value of the electric field of a line source (out of the page) placed at respective positions \(x=-5\lambda\) and \(x=+5\lambda\) in the permittivity profile shown in figure~\ref{eps_figure}(ii).  The region between the vertical dashed lines in (i) indicates the region plotted in figure~\ref{eps_figure}(ii).  Panels (iii) and (iv) are for identical parameters, but we have taken only the real and imaginary parts of the permittivity respectively.  The absence of any oscillations in panel (i) shows that the reflection is completely suppressed for incidence from the left, for all incident angles.\label{refl_figure}}
        \end{center}
    \end{figure}
    where \(\eta\) is an infinitesimal positive number.  Notice that if \(\tilde{\alpha}(k<0)=0\) then the Born approximation to the scattered field (\ref{e1}) is made up of only right--going waves, whatever the value of \(K\) (i.e. whatever the angle of incidence).  This means that to first order in \(\alpha(x)\) there is no backscattering from such a permittivity profile.  As a first order result this is not all that remarkable, but through examining all the other terms in the series (\ref{series}) we can see that there is actually no backscattering to any order.  To prove this consider the \(n^{\text{th}}\) term in the scattering series
    \begin{equation}
        e_{s}^{(n)}(x)=-k_{0}^{2}\int\frac{dk}{2\pi}\int\frac{dk'}{2\pi}G(k)\tilde{\alpha}(k-k')\tilde{e}_{s}^{(n-1)}(k')\e^{\ii k x}.\label{es_series}
    \end{equation}
    This term is also made up of only right--going waves if (i) the Fourier components of the scattered electric field \(\tilde{e}^{(n-1)}\) are zero for left--going waves \(\tilde{e}^{(n-1)}(k<0)=0\), and (ii) the Fourier components of the permittivity profile are also zero for left--going waves \(\tilde{\alpha}(k<0)=0\).  We have already established that the first term in the series (\ref{series}) is made up of entirely right--going waves when \(\tilde{\alpha}(k<0)=0\), and this argument shows that every successive term also contains only right--going waves.  There is thus zero back--scattering to all orders when \(\tilde{\alpha}(k<0)=0\).  One way to understand this result is to think that when a wave scatters multiple times from an object, for each scattering event there is a momentum change \(\Delta k\) that occurs with an amplitude proportional to \(\tilde{\alpha}(\Delta k)\).  A permittivity profile that has only positive Fourier components therefore cannot convert a right--going wave to a left--going one.
    \par
    We have established that if the permittivity (\ref{eps}) is such that the Fourier transform of its spatial dependence is zero for \(k<0\), \(\tilde{\alpha}(k<0)=0\) then a wave incident from the left onto such a medium does not give rise to any reflection, whatever the angle of incidence.  It might appear that this argument relies on a smallness condition for \(\alpha(x)\), but in the Supplementary Material we give an alternative argument that does not rely on a series expansion of the electric field, as well as deriving two exact solutions for propagation in such profiles that confirm the effect.  The Supplementary Material also contains a numerical investigation to show that an order of magnitude increase in \(\alpha(x)\) does not disturb the non--reflecting behaviour (this demonstration also shows that the real part of the permittivity can become negative and remain non--reflecting).
    \par
    In light of these properties, such non--reflecting permittivity profiles can be generally written as
    \begin{equation}
        \epsilon(x)=\epsilon_{b}+\int_{0}^{\infty}\frac{d k}{2\pi}\tilde{\alpha}(k)\e^{\ii k x},\label{nrprofile}
    \end{equation}
    which is necessarily a complex function of position.  The spatial distribution of the reactive and dissipative parts of the material response \emph{together} completely suppress reflection.  To make use of this finding, we note that equation (\ref{nrprofile}) is the same in form as the relationship between the susceptibility in the frequency and time domains which embodies the causality principle~\cite{volume8}, and one need only make the replacements \(k\to t\) and \(x\to\omega\) in (\ref{nrprofile}) in order to recover this well known formula.  As a consequence~\cite{titchmarsh1948,lucarini}, the non--reflecting permittivity profile \(\alpha(x)\) is an analytic function in the upper half complex position plane and satisfies the Kramers--Kronig relations in \emph{space}
    \begin{equation}
        \text{Re}[\alpha(x)]=\frac{1}{\pi}\text{P}\int_{-\infty}^{\infty}\frac{\text{Im}[\alpha(s)]}{s-x}ds.\label{spatialkk}
    \end{equation}
    where `\(\text{P}\)' indicates the principal part of the integral.  Therefore if we were given some \(\text{Im}[\alpha(s)]\), say as a (square integrable) function of position, a corresponding real part can be constructed from (\ref{spatialkk}) such that the reflection from the complex susceptibility is zero.  We note that if the imaginary part of \(\alpha(x)\) is symmetric about \(x=0\), then the real part calculated from (\ref{spatialkk}) will be antisymmetric, and vice versa.  Therefore the Kramers--Kronig relations generate a whole family of permittivity profiles that exhibit PT--symmetry (\(\alpha(-x)=\alpha^*(x)\)). Likewise, we also have a whole family of non--reflecting profiles where \(\alpha(x)\) exhibits PT--\emph{antisymmetry} (\(\alpha(-x)=-\alpha^*(x)\)),
a property that has already been associated with zero back--scattering in optics~\cite{ge}, just as PT--symmetry~\cite{feng}.  Actually, even purely lossy periodic media can be engineered~\cite{wu2014} such that their Bragg reflection from one side vanishes when the real and imaginary parts of their susceptibility are spatially out of phase, which is a characteristic property of Hilbert transform pairs.  However, our findings are more general than these known results, as they are also compatible with non--reflecting profiles exhibiting no definite PT--symmetry at all.
    \par
    As an initial illustration of this finding, we consider the simplest example: a permittivity profile with a single pole in the lower half position plane
    \begin{equation}
        \epsilon(x)=\epsilon_{b}+A\frac{\ii-x/\xi}{1+(x/\xi)^2}\label{eps_example}
    \end{equation}
    where \(\xi\) sets the spatial scale of the profile, and \(A\) the amplitude.  Equation (\ref{eps_example}) is plotted in figure~\ref{eps_figure} and takes a form which would be very familiar if the \(x\)--axis represented frequency rather than space.  The non--reflecting behaviour is demonstrated in figure~\ref{refl_figure} which shows the absolute value of the electric field for a point source (a line source in 2D) placed either side of \(x=0\), and compares the behaviour of the full profile (\ref{eps_example}), versus its real and imaginary parts separately.
    \par
We note that, as is well known, the Helmholtz equation (\ref{1dwv}) is equivalent to a Schr\"odinger equation in which  $-\alpha(x)$ plays the role of a potential profile. Thus, the spatial Kramers--Kronig relations generate a large family of \emph{complex} non-reflecting potential profiles. Needless to say, as the relation given by Eq.(\ref{nrprofile}) is a sufficient, but not necessary condition, real non--reflecting profiles also exist which are perfectly transparent (for example, as mentioned in the introduction the potential \(V(x)=U_{0}\,\text{sech}^{2}(x/a)\) is known to be non-reflecting for quantum particles when \(U_{0}\) takes specific values~\cite{volume3,lekner2007}).
    \par
\begin{figure}[h!]
        \begin{center}
	\includegraphics[width=10cm]{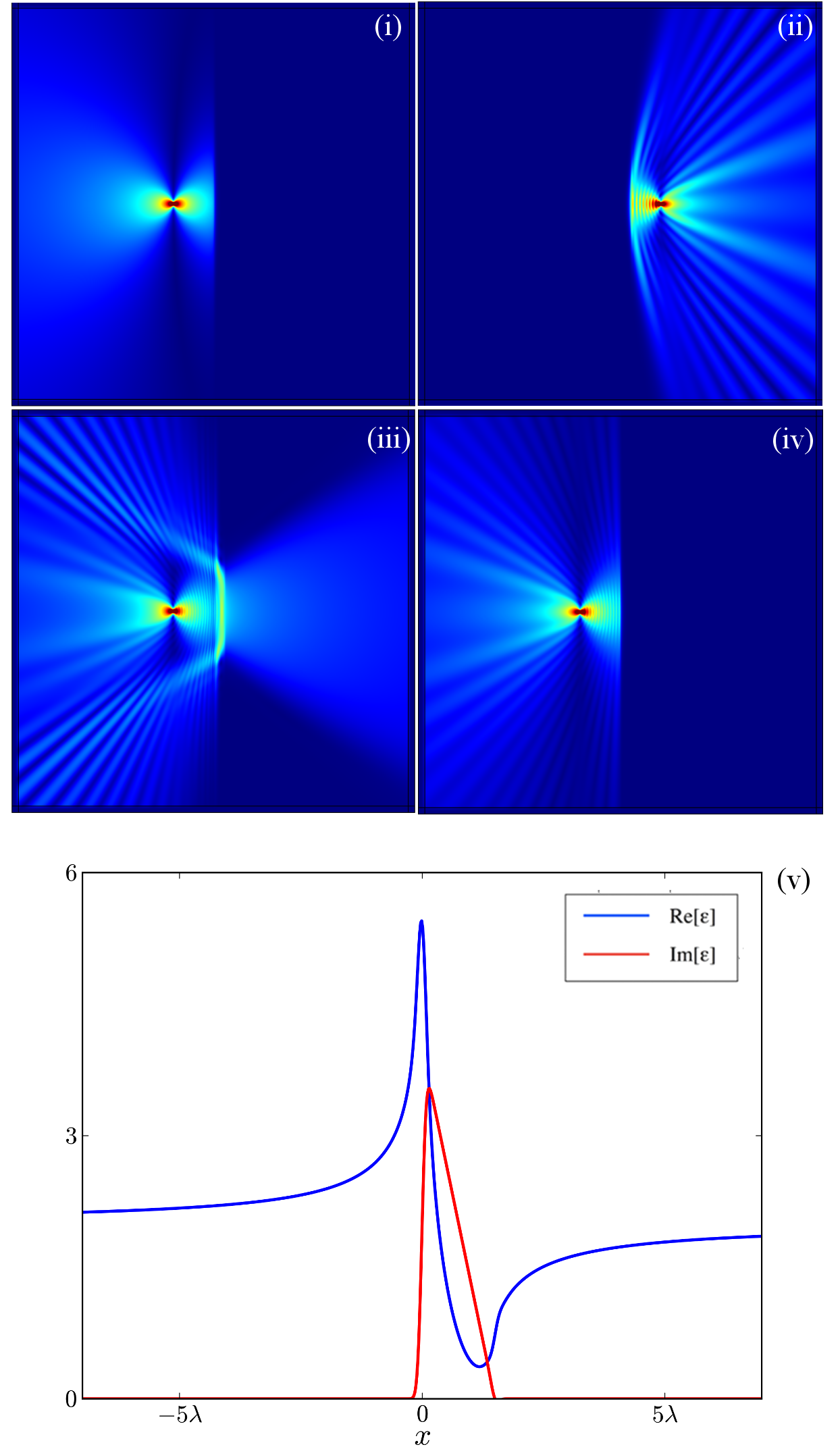}
        \caption{Panel (i) shows the absolute value of the field of a dipole source aligned along the \(y\) axis, and placed at \(x=-5\lambda\) in front of the permittivity profile shown in panel (iii).  Panel (ii) the same situation, but with the dipole placed at \(x=5\lambda\).  Panels (iii) and (iv) show the behaviour when we retain only the real and imaginary parts of \(\alpha(x)\) respectively.  The imaginary part of the permittivity profile is given by (\ref{imeps}) with \(\epsilon_{b}=2\), \(L=1.5\lambda\), \(\xi=0.1\lambda\), and \(h=1.0\).  The real part of the profile was numerically calculated using (\ref{spatialkk}).  The scale on the axes of the field plots is the same as in figure~\ref{refl_figure}\label{bragg_figure}.}
        \end{center}
    \end{figure}

    The above analysis was carried out for TE polarization, but from our argument in terms of multiple scattering one might expect that these profiles are also non--reflecting for TM polarized waves.  We now show that this is the case, given certain additional conditions on the permittivity.  The TM polarization obeys the equation
    \begin{equation}
        \boldsymbol{\nabla}\boldsymbol{\cdot}[\epsilon^{-1}(x)\boldsymbol{\nabla}H_{z}]+k_{0}^{2}H_{z}=0.\label{tmwave}
    \end{equation}
    In analogy with the foregoing argument, we write \(\epsilon^{-1}(x)=\epsilon_{b}^{-1}+\beta(x)\) and \(H_{z}=h_{z}(x)e^{i k_{y}y}\).  Equation (\ref{tmwave}) then takes the form
    \begin{equation}
        \frac{d^{2}h_{z}(x)}{dx^{2}}+K^{2}h_{z}(x)=-\epsilon_{b}\left[\frac{d}{dx}\left(\beta(x)\frac{dh_{z}(x)}{dx}\right)-k_{y}^{2}\beta(x)h_{z}(x)\right]
    \end{equation}
    Comparing the above equation with that governing TE polarized waves (\ref{1dwv}) it can be shown that the equivalent of (\ref{es_series}) is given by
    \begin{equation}
        h_{s}^{(n)}(x)=\epsilon_{b}\int\frac{dk}{2\pi}\int\frac{dk'}{2\pi}G(k)\tilde{\beta}(k-k')(k_{y}^{2}+kk')\tilde{h}_{s}^{(n-1)}(k')e^{ikx}
    \end{equation}
    Therefore if \(\tilde{\beta}(k<0)=0\) then reflection of the TM polarization is also suppressed.  However, it is not necessarily the case that both \(\tilde{\alpha}(k<0)=0\) and \(\tilde{\beta}(k<0)=0\).  For both equations to hold simultaneously we need both \(\epsilon(x)-\epsilon_{b}\) and \(\epsilon^{-1}(x)-\epsilon_{b}^{-1}\) to be analytic functions in the upper half complex position plane.  In particular,  if \(\epsilon(x)\) satisfies the spatial Kramers-Kronig relations it will be free of zeros in the upper half plane when \(\text{Im}[\epsilon(x)]\) takes only one sign along the real axis (a proof of this property of analytic functions can be found in~\cite{volume5}).  Therefore lossy media obeying the Kramers--Kronig relations in space will not reflect radiation of either polarization for any angle of incidence, which establishes the generality of the finding illustrated in figure~\ref{bragg_figure}.  Meanwhile for profiles exhibiting a combination of loss and gain where there are zeros in the upper half position plane will suppress reflection for only one of the two polarizations.  Note that because the TM polarization is sensitive to zeros in \(\epsilon(x)\) then, unlike the TE polarization, the non--reflecting behaviour is sensitive to to the value of \(\epsilon_{b}\).
    \par  
 As an example that demonstrates the generality of our finding we now take a permittivity profile with an imaginary part given by
    \begin{equation}
        \text{Im}[\epsilon(x)]=\frac{h(L-x)}{L}[1+\text{erf}(x/\xi)][1+\text{erf}((L-x)/\xi)]\label{imeps}
    \end{equation}
    which represents a smoothed triangle function, where `\(\text{erf}\)' is an error function, \(h\) is the height, \(L\) the length, and \(\xi\) characterizes the smoothness of the corners.  Numerically calculating the integral (\ref{spatialkk}) we obtain the real part of the permittivity that, when added to \(i\) times (\ref{imeps}) is necessary to reduce the reflection to zero.  The full function is shown in figure \ref{bragg_figure}(v), and unlike (\ref{eps_example}) has no definite parity symmetry.  Figures~\ref{bragg_figure}(i--iv) then show that the resulting profile is non--reflecting for TM polarized waves incident from the left, but that there is reflection from the right and for the real and imaginary parts of \(\alpha(x)\) taken separately.
    \par
    A general result in the theory of reflection~\cite{lekner1987} is that when waves are incident onto a generic planar interface, at angles close to grazing (\(k_{y}\sim k_{0}\)) the reflectivity usually approaches unity.  The above findings at first sight appear to contradict this result.  There are some other somewhat surprising features of these profiles which are related to this.  For instance if we could construct a medium with a permittivity profile that obeys the spatial Kramers--Kronig relations over all frequencies (this property does not obviously contradict the Kramers--Kronig relations in frequency), then it would be non--reflecting for all angles of incidence and all frequencies.
    
	This surprising behaviour stems from the fact that strictly speaking the profiles we have calculated are of infinite extent so that there is no `interface' to speak of, and no natural length scale associated with the profiles.   The importance of the infinite extent of the profiles is indicated by the presence of the long tails evident in the real parts of \(\epsilon(x)\) shown in figures~\ref{eps_figure} and \ref{bragg_figure}.  In practice we must confine these infinite profiles to a finite region of space through truncating these tails, which can be achieved through multiplying the profile by an envelope function \(U(x)\)
    \begin{align*}
        &\alpha'(x)=\alpha(x)U(x)\\
        \to\;&\tilde{\alpha}'(k)=\int_{0}^{\infty}\frac{dk'}{2\pi}\tilde{U}(k'-k)\tilde{\alpha}(k')
    \end{align*}
    where \(U(x)=0\) at some distance from the centre of the profile.  This truncation naturally introduces a length scale into the permittivity profile, and through doing this we find that \(\alpha'(k<0)\neq0\).  However, if \(\tilde{U}(\Delta k)\) is sharply peaked around \(\Delta k=0\), \(\tilde{\alpha}'\) will only be non--zero for negative \(k\) of a small magnitude and nearly all of the non--reflecting behaviour can be retained.  For example if we use \(U(x)=\exp(-(x/a)^{2})\), then \(\tilde{U}(\Delta k)=(a/2\sqrt{\pi})\exp(-a^{2}(\Delta k)^{2}/4)\), which rapidly goes to zero beyond around \(\Delta k=-2/a\).  The consequence of this is that waves close to grazing will now be reflected by the profile.  The Supplementary Material contains a further discussion of this effect, where it is numerically demonstrated that it is possible to confine the profile (\ref{eps_example}) to a slab of a few wavelengths thickness, introducing reflection at close to grazing incidence, while otherwise retaining the non--reflecting behaviour.
    \par
While the non-reflecting property of specific classes of real potential profiles have long been studied~\cite{kay}, more recently the analogous behaviour of PT--symmetric complex susceptibility profiles have been considered in optics. Yet, to the best of our knowledge the simple and general relation here discussed between the one-sided absence of reflection and the analytic extension of the spatially dependent susceptibility to one half of the complex position plane has not been pointed out before.  We have shown how the corresponding Kramers-Kronig relations in space can be used to generate a large family of non-reflecting profiles as they provide a sufficient condition for being non--reflecting on one side.  If the profile is also free of zeros in the upper or lower half complex position plane (the half plane being determined by whether reflection vanishes from the left or the right) then the profile is also non--reflecting for both polarizations.  In practice the catch is that the profiles have very long tails which must be truncated, and where one chooses to perform the truncation determines the range of angles and frequencies that are not reflected.  Nevertheless, the advantage of this method is that it requires us only to be able to manipulate the real and imaginary parts of an isotropic permittivity, and is in principle valid for any wave equation, including the Schr\"odinger equation. While the Kramers-Kronig relations in the frequency domain are a cornerstone of optics, it is hoped that the spatial Kramers-Kronig relations will provide at least some guidance and insight in the development of metamaterials based on judiciously chosen susceptibility profiles.
\\
    \acknowledgments
    SARH acknowledges financial support from the EPSRC under Program Grant EP/I034548/1, and thanks Scuola Normale Superiore (Pisa) for its hospitality.  The authors would like to thank J. B. Pendry, T. G. Philbin, C. King, T. C. Constant, A. Di Falco, J. R. Sambles, E. Hendry, I. R. Hooper, A. P. Hibbins, J.-H. Wu, V. Agranovich and V. Lucarini for useful discussions.  In particular J. B. Pendry and J. R. Sambles are both to be thanked for separately pointing out the limit of grazing incidence.
    
\end{document}